\documentclass[prd,aastex,amsmath,amssymb,twocolumn,floatfix,aps,showpacs]{revtex4-2} 

\usepackage{graphicx} 
\usepackage{amsmath}
\usepackage{amssymb}
\usepackage{textcomp}
\usepackage{xcolor}
\definecolor{myblue}{rgb}{0.0, 0.0, 0.6}
\usepackage{hyperref}
\hypersetup{
  colorlinks = true,
  citecolor  = myblue,
  linkcolor  = myblue,
  urlcolor   = myblue
}

\begin{document}

\title{
  Absorptive corrections to the electromagnetic form factor in high-energy elastic proton-proton scattering
}%

\author{A.~A.~Poblaguev}
\email{poblaguev@bnl.gov}
\affiliation{
  Brookhaven National Laboratory, Upton, New York 11973, USA
}

\date{September 16, 2024}

\begin{abstract}
  Recently, it was noted that absorptive corrections to the electromagnetic form factor in high-energy proton-proton scattering are important for the theoretical interpretation of the $p^\uparrow{p}$ and $p^\uparrow{A}$ analyzing power $A_\text{N}(t)$ measurements with the Hydrogen Jet Target polarimeter (HJET) at RHIC. Here, a concise expression for the absorptive correction was derived within the eikonal approach. The resulting analysis reveals a systematic bias, nearly independent of the beam energy, in the experimental determination of the real-to-imaginary ratio $\rho$ when absorption effects are overlooked in the data analysis. Quantification of this bias, as $\rho^\text{meas}=\rho + (0.036\pm0.016)_\text{bias}$, was achieved using a Regge fit applied to available proton-proton measurements of $\rho^\text{meas}(s)$ and $\sigma^\text{meas}_\text{tot}(s)$. Considering the potential impact of such an effect on the experimentally determined $A_\text{N}(t)$, one may enhance consistency between the HJET and STAR measurements of the hadronic spin-flip amplitude. While the sign of the bias in the value of $\rho$ aligns with the anticipated effective increase in the proton charge radius in $pp$ scattering due to absorption, it amplifies the observed discrepancy between $\sigma^\text{meas}_\text{tot}$ and $\rho^\text{meas}$ values at $\sqrt{s}=13\,\text{TeV}$ as measured in the TOTEM experiment. Evaluation (using published TOTEM data) of the measured proton-proton $d\sigma/dt$ dependence on the absorptive corrections indicated that possible soft photon corrections to the hadronic amplitude slope may be essential for such data analysis.
\end{abstract}


\maketitle

\section{Introduction}

Forward elastic proton-proton scattering is notable for having an electromagnetic amplitude that can exceed the hadronic one. Because the electromagnetic amplitude is well understood, experimental measurements in the Coulomb-nuclear interference (CNI) region are valuable for probing the structure of the hadronic amplitude. However, interpreting CNI effects can be challenging due to the presence of radiative corrections\,\cite{Afanasev:2023gev}. Depending on the physical process under study, various theoretical techniques have been developed to accurately describe the experimental observables.

\begin{figure}[b]
  \begin{center}
  \includegraphics[width=\columnwidth]{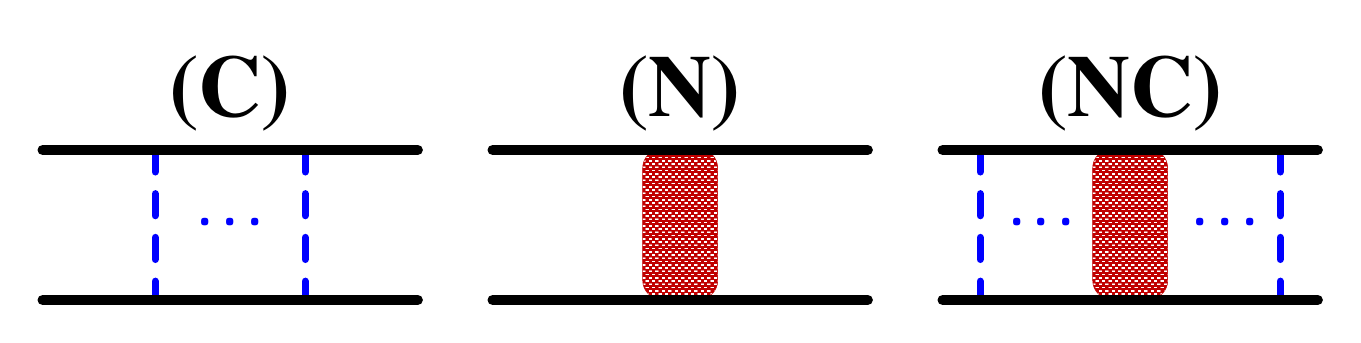}
  \end{center}
  \caption{ \label{fig:graphs}
    Three types of elastic proton-proton scattering: (C) electromagnetic, including multiphoton exchange; (N) bare hadronic; and (NC) combined hadronic and electromagnetic.
  }
\end{figure}

In $\mathit{pp}$ scattering, the exchange of multiple soft photons, as depicted schematically in Fig.\,\ref{fig:graphs}, plays a significant role. A key consequence of this multiple photon exchange is the induction of Coulomb phases, $\Phi_\text{C}^\lambda$ and $\Phi_\text{NC}^\lambda$, which affect the electromagnetic and hadronic $pp$ amplitudes, respectively. While theoretical estimates of these Coulomb phases exhibit a nonvanishing dependence on the photon mass $\lambda\to 0$, introduced to avoid infrared divergence, the phase difference $\delta_C = \Phi_\text{C}^\lambda - \Phi_\text{NC}^\lambda$ remains free of such uncertainties. The theoretical understanding of Coulomb phases in high-energy $pp$ scattering has a long history, as partially discussed in Refs.\,\cite{Akhiezer:1945,Bethe:1958zz,Solov'ev:1965,West:1968du,Cahn:1982nr}, with ongoing investigations into the theoretical aspects of Coulomb-nuclear interference.

Here, we adopt the eikonal approach, which was used in Refs.\,\cite{Kopeliovich:2021rdd,Kopeliovich:2023xtu} to emphasize the significance of absorptive corrections in interpreting experimental measurements of the transverse analyzing power $A_\text{N}(t)$ in elastic $p^{\uparrow}p$ and, particularly, $p^{\uparrow}A$ scattering. In elastic $\mathit{pp}$ scattering, the proton's electromagnetic form factor is effectively altered due to the dependence of the inelastic scattering probability on the impact parameter. As a result, both the differential cross section $d\sigma/dt$ and the analyzing power $A_\text{N}(t)$ are modified in the CNI region.

The primary goal of this paper is to obtain a compact expression for the absorptive correction which can be used in the $A_\text{N}(t)$ measurements with the Relativistic Heavy-Ion Collider hydrogen-gas jet target (HJET) polarimeter\,\cite{Poblaguev:2019saw,Poblaguev:2020qbw,Poblaguev:2023gip}. For that, the absorptive correction was related to the previously calculated $\Phi_\text{NC}^\lambda$ \cite{Kopeliovich:2000ez} using the eikonal approach. A simple method to remove the photon mass dependent part of $\Phi_\mathrm{NC}^\lambda$ was considered. The remaining ambiguity in the value of $\Phi_\mathrm{NC}$ was constrained by examining the $\rho(s)$ and $\sigma_\text{tot}(s)$ Regge fit dependence on the absorptive correction. Possible effects of the correction for the experimental measurements of forward elastic proton-proton scattering will be discussed.

By default, numerical estimates below will be given for 100\,GeV proton beam scattering off a fixed target proton.

\section{The Eikonal Model}

Following Ref.\,\cite{Kopeliovich:2000ez}, for elastic high-energy forward $pp$ scattering, the Coulomb amplitude in the Born approximation can be written as
\begin{equation}
  f_C(t) = \frac{2\alpha}{t}e^{B_Et/2},
  \label{eq:fC}
\end{equation}
where $\alpha$ is the fine structure constant, $t$ stands for the momentum transfer squared, and
\begin{equation}
  B_E = \frac{2\langle{r_E^2}\rangle}{3} = 12.1\,\text{GeV}^2
  \label{eq:B_E}
\end{equation}
is expressed via the proton charge radius \cite{Workman:2022ynf}. Applying the Fourier transformation, one obtains
\begin{align}
  \chi_C(\boldsymbol{b}) &= \frac{1}{2\pi}\int d^2\boldsymbol{q_T} e^{-i\boldsymbol{bq_T}}\,f_C(q_T^2)
  \nonumber \\
  &= -2\alpha\int_0^\infty \frac{q_Tdq_T}{q_T^2+\lambda^2}e^{-B_Eq_T^2/2}J_0(bq_T),
  \label{eq:chiC}
\end{align}
where $\boldsymbol{q_T}$ is the transverse momentum of the scattered proton ($q_T^2\!\approx\!-t$), $J_0(z)$ is the Bessel function of zero order, and a small photon mass $\lambda$ is included in Eq.\,\eqref{eq:chiC} to keep the integral finite.

Considering multiple photon exchanges, the net long-range Coulomb amplitude can be written\,\cite{Kopeliovich:2000ez} as
\begin{align}
  f_C^\gamma(t) &= \int_0^\infty bdbJ_0(bq_T)i\left[1-e^{i\chi_C(b)}\right]
  \nonumber \\
  &= f_C(t)e^{i\Phi_\text{C}^\lambda}.
  \label{eq:chiCC}
\end{align}

For the amplitude normalization used, the optical theorem relates the hadronic elastic $pp$ amplitude to the total cross section $\sigma_\text{tot}\!=\!38.39\,\text{mb}$\,\cite{Fagundes:2017iwb}, as
\begin{align}
  f_N(t) &= (i+\rho)\frac{\sigma_\text{tot}}{4\pi}e^{Bt/2}
  \nonumber \\
  &= \frac{-2\alpha(i+\rho)}{t_c}e^{Bt/2},
\end{align}
where $\rho\!=\!-0.079$\,\cite{Fagundes:2017iwb}, $B\!=\!11.2\,\text{GeV}^{-2}$\,\cite{Bartenev:1973jz}, and $-t_c\!=\!8\pi\alpha/\sigma_\text{tot}\!=\!1.86\!\times\!10^{-3}\,\text{GeV}^2$. In the eikonal analysis, the multiple photon exchange leads to
\begin{align}
  f_N^\gamma(t) &= (i+\rho)\int_0^\infty bdbJ_0(bq_T)\gamma_N(b)e^{i\chi_C(b)}
  \nonumber \\
  &= f_N(t)e^{i\Phi_\text{NC}^\lambda}, \label{eq:gammaNC} \\
  \gamma_N(b) &= \frac{-2\alpha}{Bt_c}e^{-b^2/2B}.
  \label{eq:gammaN}
\end{align}

For the Coulomb amplitude\,\eqref{eq:fC}, the phases $\Phi_\text{C}^\lambda$ and $\Phi_\text{NC}^\lambda$ were calculated analytically\,\cite{Kopeliovich:2000ez} (in lowest order of $\alpha$ and $\lambda\to 0$):
\begin{align}
  \Phi_\text{C}^\lambda/\alpha &= \ln{\frac{B_E\lambda^2}{2}} + 2\gamma + \ln w%
  -\text{Ei}\left(\frac{w}{2}\right) 
  \nonumber \\
  &+e^w\left[2E_1(w)-E_1\left(\frac{w}{2}\right)\right],
  \\
  \Phi_\text{NC}^\lambda/\alpha &= \ln\frac{B_E\lambda^2}{2} + 2\gamma + \ln{w}%
  +\ln{\frac{B^2}{B_E^2}} -\text{Ei}(z)
  \nonumber \\
  &= \ln\frac{B_E\lambda^2}{2} + \gamma + \ln\frac{B+B_E}{B_E} -z + {\cal O}(z^2)
  \label{eq:PhiNC_L}
\end{align}
where $\mathrm{Ei}$ and $E_1$ are the exponential integral functions, $w\!=\!B_Eq_T^2/2$, $z\!=\!B^2q_T^2/2(B+B_E)$, and $\gamma\!=\!0.5772\ldots$ is Euler's constant.

The phase difference, which for $t\to0$ can be approximated \cite{Cahn:1982nr} as
\begin{align}
  \delta_C(t) &= \Phi_\text{C}^\lambda\!-\!\Phi_\text{NC}^\lambda \nonumber \\
  &\approx
  -\alpha\left[\gamma+\ln{\frac{(B\!+\!B_E)|t_c|}{2}}\right] +\alpha\ln{\frac{t_c}{t}},
  \label{eq:deltaC}
\end{align}
is independent of the photon mass $\lambda$.

\section{Absorptive Corrections}

The electromagnetic form factor, as given in Eq.\,\eqref{eq:fC}, was determined in lepton-proton interactions. However, in proton-proton scattering, absorptive corrections may substantially reduce the partial elastic amplitude at small impact parameters $\boldsymbol{b}$ \cite{Krelina:2019dhm}. To evaluate this correction, the absorptive factor, $1-\gamma_N(b)$, may be considered\,\cite{Krelina:2019dhm} in the impact parameter space:
\begin{align}
  \widetilde{f}_C^\gamma(t) &= \int_0^\infty b\,db\,J_0(bq_T)i\left[1-e^{i\chi_C(b)}\right]\left[1-\gamma_N(b)\right].
  \label{eq:chiCabs}
\end{align}
Using Eqs. \eqref{eq:chiCC}, \eqref{eq:gammaN}, and \eqref{eq:gammaNC}, one readily arrives at
\begin{align}
  \widetilde{f}_C^\gamma(t) &= f_C^\gamma(t) - if_N(t)\left(1-e^{i\Phi_\text{NC}^\lambda}\right)
  \nonumber \\
  &\approx f_C(t)\left[e^{i\Phi_\text{C}^\lambda} + \Phi_\text{NC}^\lambda\,t/t_c\right],
\end{align}
which leads to the following effective correction of the electromagnetic form factor:
\begin{equation}
  B_E \to B_E^\text{eff} \approx B_E + 2\Phi_\text{NC}^\lambda/t_c.
  \label{eq:Beff}
\end{equation}

Thus, the absorption correction to the electromagnetic form factor is proportional to the Coulomb phase $\Phi_\text{NC}$, which, however, explicitly depends on the photon mass used in the calculations. On the other hand, this suggests that if Eq.\,\eqref{eq:chiCabs} adequately describes the absorption effect, there must be a way to eliminate the photon mass-dependent component from the value of $\Phi_\text{NC}$.

\section{The Coulomb Phases Dependence on the Photon Mass}

For estimates, it is helpful to consider altering $f_C^\gamma(q_T)$ and $f_N^\gamma(q_T)$ by adding a constant term to the electromagnetic Born amplitude in the impact space, $\chi_C(b) \to \chi_0 + \chi_C(b)$:
\begin{align}
  i\left[1 - e^{i\chi_C(b)}\right] &\to%
  i\left[1 - e^{i\chi_C(b)}\right]e^{i\chi_0} + i\left[1 - e^{i\chi_0}\right],
  \label{eq:chiC_}
  \\
\gamma_Ne^{i\chi_C(b)} &\to \gamma_Ne^{i\chi_C(b)}\times e^{i\chi_0}.
  \label{eq:gammaN_}
\end{align}
Notably, both $\Phi_\text{C}^\lambda$ and $\Phi_\text{NC}^\lambda$ are altered by the same value $\chi_0$. It should also be noted that in addition to phase $i\chi_0$, the right-hand side of Eq.\,\eqref{eq:chiC_} contains term $i[1 - \exp{(i\chi_0)}] \approx \chi_0$, which leads to a $\delta$-function contribution, $\chi_0\delta(\boldsymbol{q_T})$, to $f_C^\gamma(t)$. Possible appearance of such terms was discussed in Ref.\,\cite{Nekrasov:2023kxu}.

Since introducing the photon mass $\lambda$ to Eq.\,\eqref{eq:fC} switches off the Coulomb field for large values of the impact parameter ($\chi_C(b) \to 0$ if $b \gg 1/\lambda$), the photon mass-dependent constant term $\chi_0$ may be generated by a product in this procedure.

For simplicity, let $q = q_T\sqrt{B_E/2}$ and $a = \lambda\sqrt{B_E/2}$. Then,
\begin{align}
  \chi(b,a) &= \int_0^\infty \frac{dq\,qe^{-q^2}}{q^2+a^2}J_0(bq)
  = \chi_0(a) + \chi'(b,a),
  \\
  \chi_0(a) &= \chi(0,a) = \frac{e^{-a^2}E_1(a^2)}{2} \approx -\ln{a} - \gamma/2,
  \\
  \chi'(b,a) &= \int_0^\infty \frac{dq\,qe^{-q^2}}{q^2+a^2}\left[J_0(bq) - 1\right]
  \label{eq:chi'}
\end{align}

\begin{figure}[t]
  \centering
  \includegraphics[width=0.8\columnwidth]{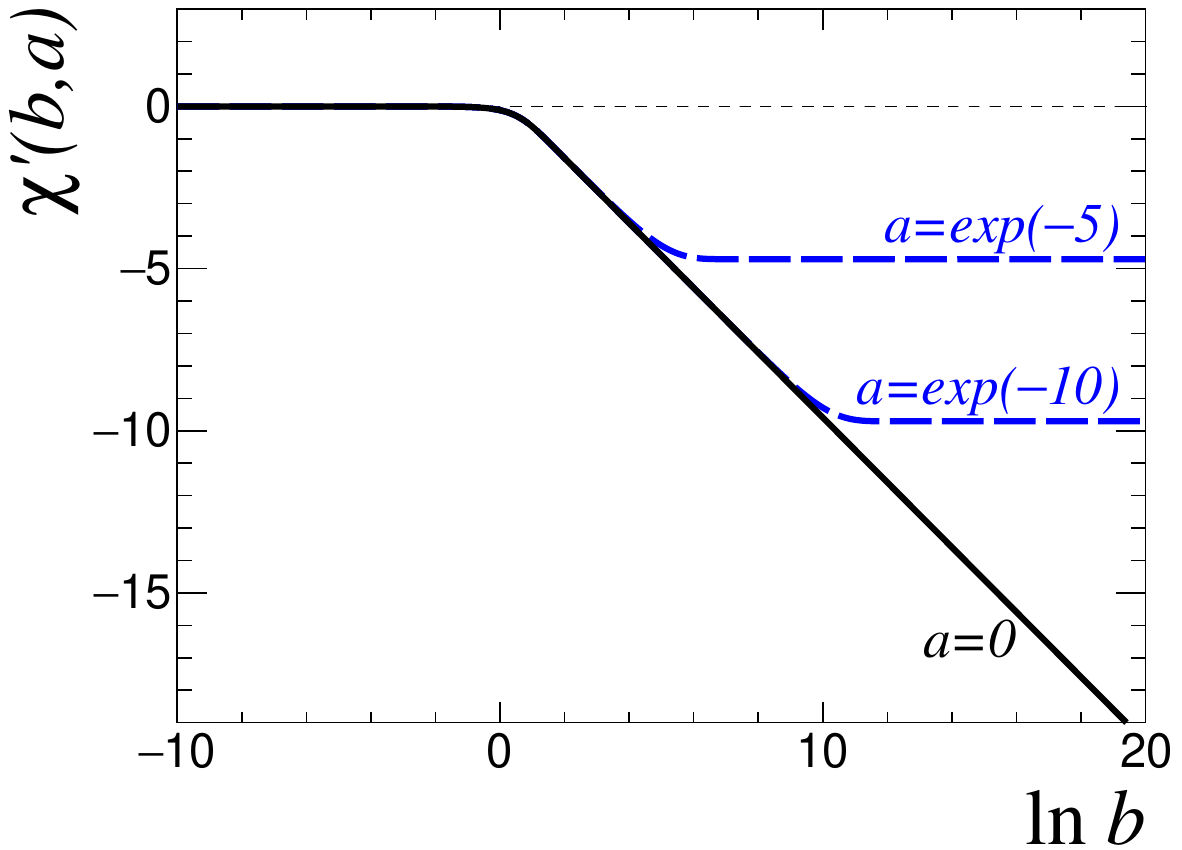}
  \caption{ \label{fig:chiC}
    Numerical calculation of $\chi'(b,a)$ [Eq.\,\eqref{eq:chi'}].
  }
\end{figure}

Numerical calculations of $\chi'(b,a)$ are depicted in Fig.\,\ref{fig:chiC}. For small values of $b$, $J_0(bq) \to 1$ and $\chi_0'(b,a) \to 0$. If $b$ is large, it was numerically found
\begin{equation}
  \chi'(b,a=0) \approx -\ln{b} + \ln{2} - \gamma/2.
\end{equation}
for $\lambda\!\to\!0$.
However, for a finite (but small) $\lambda$ and $b\!>\!1/\lambda$,
\begin{equation}
  \chi'(b,a) = \ln{a} + \gamma/2 \approx -\chi_0(a),
\end{equation}
which, after adding $\chi_0(a)$, leads to $\chi_C(b) = 0$ if $b \gtrsim 1/\lambda$. According to Eqs.\,\eqref{eq:gammaN_}, such an independent of impact parameter $b$ term $\chi_0(a)$ may be used to eliminate the Coulomb phase dependence on $\lambda$,
\begin{align}
  \Phi_\mathrm{NC} &= \Phi_\mathrm{NC}^\lambda + 2\alpha\chi_0(\lambda\sqrt{B_E/2})
  \nonumber \\
  &\approx \alpha\left[C + \ln{\frac{B+B_E}{2B_E}} + \frac{B^2t}{2(B+B_E)}\right].
  \label{eq:PhiNC}
\end{align}
However, since $\chi_0(a)$ is the value of $\chi(b,a)$ at $b = 0$ rather than a constant term $\chi_0$,
$\Phi_\mathrm{NC}$ is still determined up to some unknown constant $C$.

Substituting $\Phi_\mathrm{NC}$ into Eq.\,\eqref{eq:Beff}, one finds
\begin{align}
  B_E^\text{eff} &= \frac{2\langle r_E^2\rangle}{3} - \frac{\sigma_\text{tot}}{4\pi}\left[C + \ln{\frac{B+B_E}{2B_E}}\right].
  \label{eq:BeEff}
\end{align}
The $t$ dependent terms were omitted in Eq.\,\eqref{eq:BeEff}, first, because the $t$ dependence was not considered for $B_E$ in Eq.\,\eqref{eq:fC}, and second, such terms are small in CNI scattering and inessential beyond the CNI region (large $|t|$) where the hadronic amplitude strongly dominates.

Using the slope $B(s)$ dependence\,\cite{Bartenev:1973jz} on the center of mass squared energy $s$,  
the logarithmic term in Eq.\,\eqref{eq:BeEff} can be approximated by
\begin{equation}
  \ln{\frac{B+B_E}{2B_E}} = \ln{\left(1 + \frac{2\alpha'}{B_E}\ln{\sqrt{\frac{s}{s_E}}}\right)},
\end{equation}
where $\alpha' = 0.278 \pm 0.024\,\text{GeV}^{-2}$\,\cite{Bartenev:1973jz} and $s_E^{1/2} \approx 32\,\text{GeV}$. Even for $s^{1/2} = 13\,\text{TeV}$, such a correction to $\Phi_\mathrm{NC}$ is small, $\sim 0.002$, and this term may be considered as negligible.

\section{Systematic Bias in Measurements of the Real-to-Imaginary Ratio}

For elastic proton-proton scattering, the differential cross-section dependence on $t$ can be written as
\begin{equation}
  \frac{d\sigma}{dt} = \frac{\sigma_\text{tot}^2}{16\pi}\left[ \left(\frac{t_c}{t}\right)^2-2(\rho+\delta_C)\frac{t_c}{t}+1+\rho^2\right]e^{Bt}.
  \label{eq:dsdt}
\end{equation}
Here, for simplicity, $t_c/t$ denotes the ratio of the electromagnetic and imaginary parts of the hadronic amplitudes $(t_c/t)\exp{\left[(B_E - B)t/2\right]}$.

The absorptive correction effectively modifies the electromagnetic amplitude as
\begin{equation}
  \frac{t_c}{t}\to\frac{t_c}{t}+\alpha C +\frac{\alpha B^2t_c}{2(B+B_E)}\frac{t}{t_c}.
  \label{eq:fC-corr}
\end{equation}

The $t$-dependent term in $\Phi_\text{NC}$ results in the following alteration in $d\sigma/dt$
\begin{align}
  1+\rho^2 &\to 1+\rho^2 + \frac{\alpha B^2t_c}{B+B_E} \approx 1+\rho^2-7\times10^{-5},
\end{align}
which is much smaller than the typical uncertainties $2\Delta\sigma_\text{tot}/\sigma_\text{tot}$ due to experimental accuracy $\Delta\sigma_\text{tot}$ for the total cross-section. 

The $\alpha C$ term in \eqref{eq:fC-corr} leads to a systematic bias\,\cite{Poblaguev:2019vho}
\begin{equation}
  \rho^\text{meas} = \rho - \alpha C,
  \label{eq:rho_meas}
\end{equation}
in the experimental determination of the real-to-imaginary ratio $\rho$. Such a bias, which is expected to have only a very weak dependence on $s$, may potentially be recognized in a Regge fit of the experimental values of $\sigma_\text{tot}(s)$ and $\rho(s)$.

For unpolarized proton-proton scattering, the elastic amplitude's dependence on $s$ can be approximated (model AU--L$\gamma$=2 of Ref.\,\cite{Fagundes:2017iwb}) by
\begin{equation}
  \sigma_\text{tot}(s)\left[i+\rho(s)\right] = 
  f_PP(s) + f_+R^+(s) + f_-R^-(s),
\label{eq:Regge_nf}
\end{equation}
where Regge pole terms
\begin{equation}
  R^\pm(s,\alpha_\pm)=
  \left[1\pm e^{-i\pi\alpha_\pm}\right]\left(\frac{s}{4m_p^2}\right)^{\alpha_\pm-1}
\end{equation}
are encoded as $R^+$ for ($f_2,a_2$) and $R^-$ for ($\omega,\rho$), and a Froissaron parametrization is used for the Pomeron contribution
\begin{equation}
  P(s,\alpha_F)=\pi\alpha_F\ln{\frac{s}{4m_p^2}}
  + i\left[1+\alpha_F\ln^2{\frac{s}{4m_p^2}}\right].
\end{equation}

\begin{figure}[t]
  \centering
  \includegraphics[width=0.8\columnwidth]{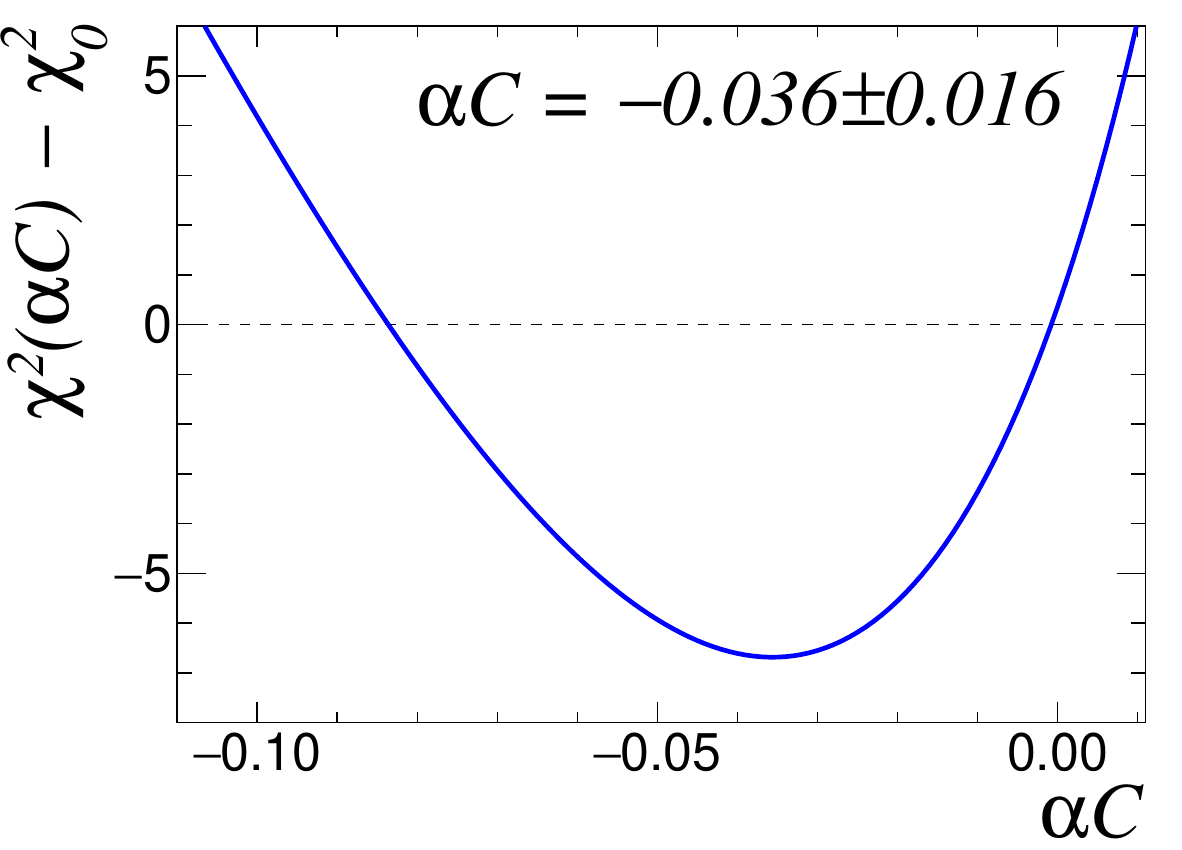}
  \caption{\label{fig:Chi2}
    Chi-squared dependence of the Regge fit of $\sigma_\text{tot}(s)$ and $\rho(s)$ on the absorptive correction parametrized by $\alpha C$. $\chi^2_0/\text{ndf}=247.2/208$ is the chi-squared calculated with no absorption.
  }
\end{figure}

To search for the bias $\alpha{C}+\ln{\left[(B+B_E)/2B_E\right]}$ in the $\rho(s)$ measurements, the $\sigma_\text{tot}(s)$ and $\rho(s)$ accelerator $pp$ dataset for $p_\text{Lab}>5\,\text{GeV}$ was taken from \cite{Zyla:2020zbs}. However, due to the known discrepancy\,\cite{TOTEM:2017sdy} between $\sigma_\text{tot}(s)$ and $\rho(s)$ measurements at $\sqrt{s}=13\,\text{TeV}$, TOTEM values of $\rho(s)$ were excluded from the dataset.

In the fit, it was found (see Fig.\,\ref{fig:Chi2})
\begin{equation}
  \alpha C = -0.036\pm0.016.
  \label{eq:alphaC}
\end{equation}
The statistical confidence of the nonzero value of the bias \eqref{eq:rho_meas} corresponds to 2.6 standard deviations. Excluding LHC (ATLAS and TOTEM) values for the total cross-section from the fit does not practically change the result, $\alpha C=-0.033\pm0.015$.

\section{Absorptive Corrections to the Measured Analyzing Power $\boldsymbol{A_\text{N}}$}

Omitting some small terms, the high-energy forward elastic $p^{\uparrow}p$ analyzing power can be expressed\,\cite{Kopeliovich:1974ee,Buttimore:1998rj} via single spin-flip $\phi_5(t)$ and nonflip $\phi_+(t)$ helicity amplitudes as
\begin{equation}
A_\text{N}(t) = \frac{\sqrt{-t}}{m_p} \times \frac%
{(\kappa_p - 2\mathrm{Im}\,r_5)\,t_c/t - 2\mathrm{Re}\,r_5}%
{(t_c/t)^2 - 2(\rho + \delta_C)\,t_c/t + 1 + \rho^2},
\label{eq:A_N}
\end{equation}
where $\kappa_p \!=\! 1.793$ is the anomalous magnetic moment of the proton, and $|r_5| \!\sim\! 0.02$\,\cite{Poblaguev:2019saw} is the hadronic spin-flip amplitude parameter.

If the determination of $r_5$ is the main goal of the experimental data analysis in the CNI region, typically, the real-to-imaginary ratio $\rho$ is taken from the Regge fit of $\rho(s)$ and $\sigma_\text{tot}(s)$, and $\delta_C$ is calculated using Eq.\,\eqref{eq:A_N}. Since such a defined value of $\rho$ already includes the absorptive correction, one can expect\,\cite{Poblaguev:2019vho} that the absorption is already accounted for in the experimentally determined value of $r_5$. Nevertheless, since the absorption-related systematic bias modifies the Regge fit, the corresponding correction $\rho\to\rho\!+\!\Delta^\text{abs}\rho$ should be applied in Eq.\,\eqref{eq:A_N}.

For the HJET energies, $\Delta^\text{abs}\rho\!=\!-0.0041$ ($\sqrt{s}\!=\!13.76\,\text{GeV}$) and $\Delta^\text{abs}\rho\!=\!-0.0010$ ($\sqrt{s}\!=\!21.92\,\text{GeV}$), which leads\,\cite{Poblaguev:2019saw} to a small (compared to the HJET experimental accuracy) correction
\begin{equation}
\Delta r_5= (-0.11+i\,0.86)\times\Delta^\text{abs}\rho
\end{equation}
in the value of $r_5$. For the STAR measurement at $\sqrt{s}\!=\!200\,\text{GeV}$\,\cite{STAR:2012fiw}, $\Delta^\text{abs}\rho\!=\!0.0256$, and the corresponding corrections $\Delta^\text{abs}\mathrm{Re}\,r_5\!\approx\!-0.003$ and $\Delta^\text{abs}\mathrm{Im}\,r_5\!\approx\!0.022$ may noticeably improve consistency between the extrapolation of the HJET results to $\sqrt{s} \!=\! 200\,\text{GeV}$ and the STAR value (see Fig.\,3 in Ref.\,\cite{Poblaguev:2023gip}).

A larger effect of the absorptive corrections is expected in the measurement of the $p^\uparrow\mathrm{Au}$ analyzing power. It was demonstrated in Ref.\,\cite{Kopeliovich:2023xtu} that absorptive correction drastically changes the $A_\text{N}^{pAu}(t)$ dependence on $t$, allowing qualitative agreement of the theoretical calculations with the results of the measurements done at HJET\,\cite{Poblaguev:2023gip,Poblaguev:2022xoa}. However, there is still room for improving the quantitative consistency. Since the absorptive correction calculations in Ref.\,\cite{Kopeliovich:2023xtu} had been done assuming cancellation of the photon mass term, variation of the absorption contribution, e.g., by scanning the parameter $\alpha C$, may be suggested.

\section{Discussion}

Following the approach in Ref.\,\cite{Kopeliovich:2023xtu} for the absorption effect in forward elastic proton-proton scattering, it was shown that the absorptive correction to the electromagnetic form factor effectively biases the experimentally determined values of the real-to-imaginary ratio $\rho(s)$. This bias is almost independent of the center-of-mass energy $s$. An indication of such a bias \eqref{eq:alphaC} was found in the Regge fit \eqref{eq:Regge_nf} of the available proton-proton $\rho(s)$ and $\sigma_\text{tot}(s)$ measurements.

Nevertheless, it has not been definitively proven that the bias can be attributed to the absorptive correction. Therefore, alternative evaluations of the absorption are needed.

According to Ref.\,\cite{Kopeliovich:2023xtu}, the proton-nucleus analyzing powers $A_N^{p\text{Au}}(t)$, measured at HJET \cite{Poblaguev:2023gip}, clearly indicate the presence of absorption. However, in \cite{Kopeliovich:2023xtu}, the difference between the calculated (including absorption) and measured $A_N^{p\text{Au}}(t)$ was attributed to the hadronic spin-flip amplitude. Since this amplitude was not accurately predetermined, the normalization of the absorption effect remained essentially ambiguous. Nevertheless, relating the $pA$ hadronic spin-flip amplitude to the $\mathit{pp}$ amplitude \cite{Kopeliovich:2000kz,Poblaguev:2023gfl} may allow one to reliably isolate the effect of the absorptive correction in the $p^\uparrow{A}$ data analysis.

Notably, the absorptive correction to the electromagnetic form factor dependence on the photon mass $\lambda$ can be canceled if the phase $\Phi_\text{NC}^\lambda(t,\lambda)$ in Eq.\,\eqref{eq:Beff} is replaced by
\begin{equation}
\Phi_\text{NC}^\lambda(t,\lambda) \to \Phi_\text{NC}^\lambda(t,\lambda) - \Phi_\text{C}^\lambda(t,\lambda) = -\delta_C(t).
\label{eq:PhiNCmod}
\end{equation}
Consequently, the Coulomb phase in Eq.\,\eqref{eq:dsdt} will be effectively doubled,
\begin{equation}
\rho + \delta_C(t) \to \rho + p \delta_C(t), \qquad p = 2,
\end{equation}
or, equivalently,
\begin{equation}
\alpha\left[C + \ln\frac{B + B_E}{2B_E}\right] \to
-\delta_C(t_c) - \alpha \ln\frac{t_c}{t}
\approx -0.021.
\label{eq:alphaC_1}
\end{equation}
Such an estimated value of $\alpha C$ is in agreement with that given in Eq.\,\eqref{eq:alphaC}.

\begin{figure}[t]
\centering
\includegraphics[width=0.8\columnwidth]{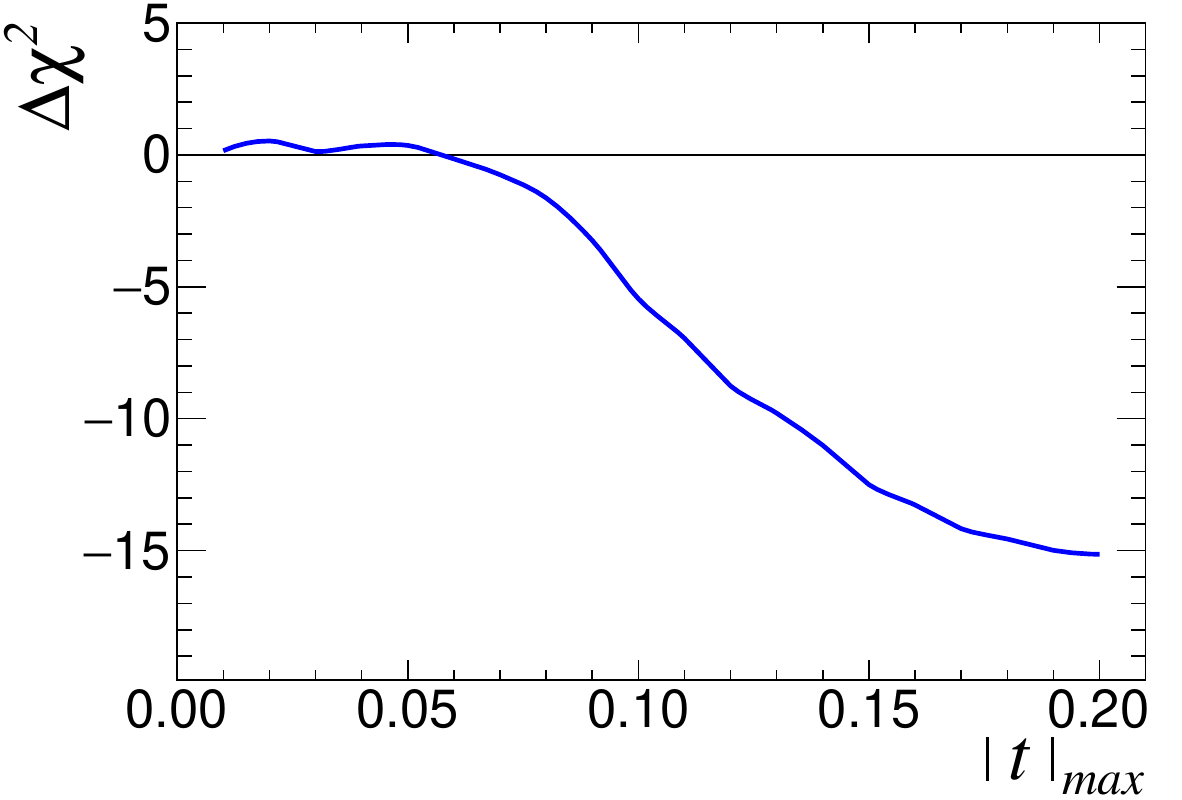}
\caption{ \label{fig:DeltaChi2}
Chi-squared difference for the TOTEM $d\sigma/dt$ data fit with ($p\!=\!2$) and without ($p\!=\!1$) absorptive correction. $|t|_\mathit{max}$ is the upper limit for the $|t|$ range in the fit.
}
\end{figure}

To evaluate the effect of the $\ln{t}$ term in Eq.\,\eqref{eq:alphaC_1} on the determination of $\rho$ from the experimental $d\sigma/dt$ distribution, the TOTEM data (Table 3 in \cite{TOTEM:2017sdy}) was used. For the $d\sigma/dt$ parametrization \eqref{eq:dsdt}, $t\!<\!|t|_\mathit{max}\!=\!0.07\,\text{GeV}$, and no absorptive corrections, the fit results are well consistent with those given in Table 4 \cite{TOTEM:2017sdy} for $N_b\!=\!1$. The $\chi^2$ difference for fits with ($p\!=\!2$) and without ($p\!=\!1$) the absorptive correction\,\eqref{eq:alphaC_1} as a function of $|t|_\mathit{max}$ is shown in Fig.\,\ref{fig:DeltaChi2}. For $|t|_\mathit{max}\!<\!0.07\,\text{GeV}$, i.e., for the CNI scattering, the logarithmic term does not degrade the quality of the fit.

Thus, the assumption \eqref{eq:PhiNCmod} is not in disagreement with available experimental data. However, to utilize such an approach, it is necessary either to correspondingly revise Eq.\,\eqref{eq:chiCabs} or, for example, to postulate (and justify) that within the applicability of Eq.\,\eqref{eq:chiCabs}, the following should be assumed:
\begin{equation}
\lim_{\lambda \to 0} \Phi_\text{C}^\lambda(t,\lambda) = 0, \qquad
\lim_{\lambda \to 0} \Phi_\text{NC}^\lambda(t,\lambda) = -\delta_C(t).
\end{equation}

In Fig.\,\ref{fig:DeltaChi2}, the significant reduction of $\chi^2$ by the $\ln{t}$ term for $|t|_\mathit{max} > 0.07\,\text{GeV}^2$ may be attributed to a possible dependence of the slope
\begin{equation}
B \to B \times \left(1 + \beta \ln{t_c/t}\right)
\label{eq:Bpar}
\end{equation}
on $t$. In the $|t|_\mathit{max} = 0.15\,\text{GeV}^2$ fit, it was found $\beta = 0.021 \pm 0.002$ ($p = 1$) and $\beta = 0.020 \pm 0.002$ ($p = 2$).

In the no absorptive correction fit, introducing $\beta$ reduced $\chi^2/\text{ndf}$ from 236.9/115 to a statistically excellent value of 105.9/114.

Comparing values of $\rho$ and $\sigma_\text{tot}$ determined in $|t| < 0.07\,\text{GeV}^2$ fit with fixed $\beta = 0.021$ and $\beta = 0$, one finds the following corrections due to the nonzero $\beta$
\begin{equation}
  \Delta^\beta\rho = +0.034, \qquad \Delta^\beta\sigma_\text{tot} = -3.4\,\text{mb},
  \label{eq:Delta_beta}
\end{equation}
which technically eliminate the observed discrepancy \cite{TOTEM:2017sdy} between measured values of $\rho$ and $\sigma_\text{tot}$ at TOTEM. However, after applying the absorptive correction \eqref{eq:alphaC_1}, the conclusion is less definitive, $\Delta^\beta\rho = +0.013$ and $\Delta^\beta\sigma_\text{tot} = -4.2\,\text{mb}$. It should also be noted that the value of $\beta$ found in the fit strongly (up to a factor of 2) depends on the $t$-range used, which suggests that the effective $B(t)$ may contain other contributions, for example, due to systematic errors in the measurement. Therefore, the result \eqref{eq:Delta_beta} should be interpreted with caution.

In the analysis \cite{TOTEM:2017sdy}, a polynomial parametrization of the slope
\begin{equation}
B(t) = \sum_{n=1}^{N_b} b_n t^{n-1}, \quad N_b = 1, 2, 3,
\label{eq:Bpol}
\end{equation}
was utilized, allowing (if $N_b > 1$) good approximation of the measured $d\sigma/dt$ beyond the CNI region, $|t| > 0.07\,\text{GeV}^2$.

Although consideration of the $B(t)$ dependence on $t$ is beyond the scope of this paper, it may be noted that Eq.\,\eqref{eq:Bpar} provides an example of a parametrization that is well consistent with the experimental $d\sigma/dt$ for large $|t|$ but leads to significant, compared to the polynomial parametrization \eqref{eq:Bpol}, changes in $\rho$ and $\sigma_\text{tot}$ in the fit.

The numerical value $\beta \approx 3\alpha$ of the correction \eqref{eq:Bpar}, as well as its logarithmic dependence on $t$, suggest it may be caused by a possible radiative correction to the slope $B$. If so, calculating the $\ln{t_c/t}$ corrections to the hadronic slope $B(t)$ may be critically important for the experimental study of forward elastic proton-proton scattering. Depending on the value of $\beta$ evaluated, a revision of the previously measured real-to-imaginary ratios $\rho(s)$ may be required.

\section*{Acknowledgments}

The author acknowledges support from the Office of Nuclear Physics in the Office of Science of the US Department of Energy. This work is authored by employee of Brookhaven Science Associates, LLC under Contract No. DE-SC0012704 with the U.S. Department of Energy.

%

\end{document}